\PassOptionsToPackage{table}{xcolor}
\documentclass[pdflatex,sn-mathphys-ay]{sn-jnl}

\usepackage{graphicx}
\usepackage{amsmath,amssymb,amsfonts}
\usepackage{array}
\usepackage{booktabs}
\usepackage{multirow}
\usepackage{float}
\usepackage[title]{appendix}
\usepackage{enumitem}
\setlist{nosep,leftmargin=1.4em}

\makeatletter
\AtBeginDocument{\immediate\write\@auxout{\string\citation{snsettings}}}
\makeatother

\newcommand{\Fmax}{F_{\max}}
\newcommand{\KaHIII}{KaH-III}
\newcommand{\KaHR}{KaH-robust}
\newcommand{\pkg}[1]{\textsf{#1}}

\newcommand{\ubm}[1]{\mbox{\unboldmath$#1$}}
\usepackage{tikz}
\definecolor{orcidgreen}{HTML}{A6CE39}
\gdef\orcidlogo{\raisebox{-0.2ex}{\tikz[baseline=-0.6ex]{\fill[orcidgreen] (0,0) circle (0.75ex);
  \node[white, inner sep=0pt, font=\sffamily\bfseries\fontsize{3.6}{3.6}\selectfont] at (0,0) {iD};}}}

\theoremstyle{thmstyleone}
\newtheorem{theorem}{Theorem}
\newtheorem{proposition}{Proposition}
\theoremstyle{thmstyletwo}

\begin{document}

\title[Three-group tests for heteroscedasticity]{Three-group variance-ratio tests for heteroscedasticity in linear regression: an exact null distribution and an outlier-resistant version}

\author[1]{\fnm{Ahmed} \sur{El-Kotory}\,\orcid{https://orcid.org/0009-0000-1485-0768}}\email{ahmed.elkatory@alexu.edu.eg}
\author*[1]{\fnm{Ebrahim Khaled} \sur{Ebrahim}\,\orcid{https://orcid.org/0009-0006-7839-8778}}\email{ebrahimkhaled@alexu.edu.eg}
\affil[1]{\orgdiv{Department of Applied Statistics}, \orgname{Faculty of Business, Alexandria University}, \orgaddress{\city{Alexandria}, \country{Egypt}}}

\abstract{Tests for heteroscedasticity in linear regression lose their level or power in three situations: when the data contain outliers, when the variance is not monotone, and when it changes along a variable outside the regressors, such as time. The proposed tests sort the observations by any chosen variable, split them into three equal parts, fit the regression in each part and compare the largest with the smallest error scale. With least squares fits and normal errors, the ratio follows Hartley's maximum \ubm{F} distribution with three groups exactly. With least trimmed squares fits, the ratio resists outliers spread along the ordering. Its square is approximately a maximum \ubm{F} ratio with effective degrees of freedom, whose limit we derive in closed form from the influence function of the trimmed variance. A factorial simulation covered 72 settings of variance shape, ordering variable, number of regressors, contamination and sample size. The robust test had the highest mean size-adjusted power, 67\%, against 45\% for the next test; when the variance changed along a regressor, which all tests were given, it led White's test after an outlier screen, 67\% against 56\%. With heavy-tailed \ubm{t_3} errors its lead was similar, 64\% against 35\%. Unlike the tests built on the regressors, it can follow a variance changing along time, and it identifies the part of the data where the variance changes. Two published robust versions of the Goldfeld--Quandt test, as implemented from their descriptions, did not hold their nominal level. The R package \pkg{KOTORY} implements the methods.}



\maketitle

%======================================================================
\section{Introduction}\label{sec:intro}

Least squares estimates of a linear regression remain unbiased when the error variance changes across observations, but their standard errors, confidence intervals and tests do not. Testing for heteroscedasticity is therefore a routine step of regression diagnostics. The literature offers many tests: the ordered-sample test of \citet{GoldfeldQuandt1965} and its generalization to a class of ordered-sample tests by \citet{Szroeter1978}, the score tests of \citet{BreuschPagan1979} and \citet{Koenker1981}, the general test of \citet{White1980}, the BAMSET test of \citet{Ramsey1969}, the point-optimal test of \citet{EvansKing1988}, tests for transformation models \citep{HuskovaMeintanisPretorius2022}, a nonparametric test for multiple regression \citep{ZambomKim2017} and many others; \citet{Farrar2025review} review and compare them. A detected change in variance also changes how the model should be fitted, for example by quantile methods that pool information across quantiles \citep{KimJung2025}.

Almost all of these tests are built on least squares residuals, and least squares residuals are distorted by outliers. The consequence works in both directions. A few gross errors in a homoscedastic sample inflate the residual variance of the region where they fall, so the test reports heteroscedasticity that is not there. A few gross errors in the low-variance region of a heteroscedastic sample raise its variance to the level of the rest, so the test misses heteroscedasticity that is there \citep{RanaMidiImon2008,AlihOng2015,BerenguerRicoWilms2021}. Tests for equal variances are, moreover, sensitive to non-normal errors in general \citep{Box1953}. Once the data are split into groups, the problem becomes a comparison of scales across samples, for which the Levene, Brown--Forsythe and Fligner--Killeen tests are the robust standards \citep{BrownForsythe1974,FlignerKilleen1976,ConoverJohnsonJohnson1981,LimLoh1996}; applied to regression residuals, they have no reference distribution that accounts for the fit and no breakdown guarantee.

Robust estimation of heteroscedastic regression models is well developed; \citet{AmadoBiancoBoente2025} review it and show that outliers are harder to handle when heteroscedasticity is present. Robust testing for heteroscedasticity has received less attention. Robust tests have been proposed along four lines. The earliest replaces least squares residuals by robust residuals, or by bounded functions of them, in a score-type test \citep{Bickel1978,CarrollRuppert1981}; the regression of absolute residuals of \citet{Glejser1969} was made robust to asymmetric and heavy-tailed errors by \citet{Im2000} and \citet{MachadoSantosSilva2000}. The second replaces least squares by a robust fit inside a classical test: \citet{RanaMidiImon2008} modified the Goldfeld--Quandt test with least trimmed squares (LTS) and median squared deletion residuals, and \citet{AlihOng2015} with a forward-search outlier screen and prediction residuals; an earlier rank-based way to resist gross errors in ordered-sample testing is the peak test of \citet{Horn1981}. The third removes outliers and applies a classical test to the rest, with theory given by \citet{BerenguerRicoWilms2021}. The fourth uses quantile regression \citep{KoenkerBassett1982,WilcoxKeselman2006}. For the two modified Goldfeld--Quandt tests, the null distribution of the modified statistic is taken from the unmodified one, which we show below can leave the level far from nominal; the outlier-removal test is asymptotically valid under symmetric errors but can be distorted by some types of contamination \citep{BerenguerRicoWilms2021}. None of these tests says where in the data the variance changes. Robust and classical tests also share two blind spots. The ordered-sample and point-optimal tests look for a variance that moves in one direction, so they miss a variance that is high at both ends and low in the middle, or the reverse. The score tests, and the robust tests of the first line, model the variance through the regressors, so they cannot follow a variance that changes over time or along another variable outside the model.

We study a simple alternative that stays close to the Goldfeld--Quandt idea and addresses the three situations above: outliers, a variance that is not monotone, and a variance that changes along a variable outside the model. The observations are sorted by a regressor or by any other variable and split into three equal parts. The regression is fitted separately in each part, and the statistic is the ratio of the largest to the smallest error scale of the parts. With least squares fits we call it \KaHIII{}; with LTS fits, \KaHR{}. The name KaH is formed from the initials of Ahmed El-Kotory, who proposed the tests (K and a), and H for heteroscedasticity; the Roman numeral III marks the three parts. Splitting ordered residuals into three groups goes back to the BAMSET test of \citet{Ramsey1969}, which applies Bartlett's test to three groups of residuals from a single fit.

Our contributions are the following.
\begin{enumerate}
  \item As for the Goldfeld--Quandt ratio, fitting the parts separately makes \KaHIII{} follow Hartley's maximum $F$ distribution \citep{Hartley1950} with three groups exactly under normal errors, for every sample size, so that its critical values and $p$-values follow from one numerical integral without simulation (Theorem~\ref{thm:exact}).
  \item For \KaHR{} we propose the approximation $T_2^2 \approx \Fmax(3,\nu^*)$ with effective degrees of freedom $\nu^*$, and derive the limit $r(\alpha)$ of $\nu^*/\nu$ in closed form for the LTS scale from the influence function of the trimmed variance (Proposition~\ref{prop:nustar}); finite-sample values are tabulated.
  \item We give the level breakdown point of \KaHR{}, a lower bound on its power breakdown point, and a residual-bootstrap reference based on the LTS fits.
  \item In a factorial comparison with the classical tests and the robust tests of \citet{RanaMidiImon2008}, \citet{AlihOng2015}, \citet{BerenguerRicoWilms2021} and \citet{WilcoxKeselman2006}, over variance shapes, ordering variables, dimensions and contamination, \KaHR{} has the highest mean size-adjusted power by a wide margin (Section~\ref{sec:factorial}), and it keeps that lead under heavy-tailed errors (Section~\ref{sec:heavy}). On two of the four benchmark data sets used by the authors of the competing robust tests (housing and savings) it detects the heteroscedasticity that their planted outliers hide from the classical tests.
  \item The methods are implemented in the R package \pkg{KOTORY} \citep{KOTORY}.
\end{enumerate}

Section~\ref{sec:method} defines the tests and gives their null distributions. Section~\ref{sec:sim} reports the simulation study and Section~\ref{sec:data} the examples. Section~\ref{sec:software} describes the software, and Section~\ref{sec:discussion} discusses limitations. Proofs are in the Appendix; the table of $\nu^*$, the implementation of the competing tests and the full RGQ study are in Online Resource~1.

%======================================================================
\section{The three-group tests}\label{sec:method}

\subsection{Model and splitting}\label{sec:split}

Consider the linear model $y_i = \beta_0 + \mathbf{x}_i^\top \boldsymbol\beta + \varepsilon_i$, $i=1,\dots,n$, with $p$ regressors, a fixed design and independent errors with $E(\varepsilon_i)=0$ and $\operatorname{var}(\varepsilon_i)=\sigma_i^2$. The null hypothesis is $H_0\colon \sigma_1^2=\dots=\sigma_n^2=\sigma^2$. The alternative of interest is a variance that changes along a sorting variable $z$, which is usually one of the regressors and may be any variable that does not depend on the errors.

Sort the observations by $z$ and let $m=\lfloor n/3 \rfloor$. The three parts are
\[
P_1=\{1,\dots,m\},\quad P_2=\{\lfloor n/3\rceil+1,\dots,\lfloor 2n/3\rfloor\},\quad P_3=\{\lceil 2n/3+1\rceil,\dots,n\},
\]
where $\lfloor\cdot\rceil$ denotes rounding to the nearest integer. Each part contains $m$ observations; when $n$ is not a multiple of three, the one or two observations at the boundaries between the parts are left out. The regression, with an intercept, is fitted separately in each part, which requires $m>p+1$. We use three parts because two cannot detect a variance that is high or low in the middle, while more parts leave fewer observations per part, lower the breakdown count of each part (Section~\ref{sec:breakdown}) and raise the critical value; the reference distribution below and the software allow any number of parts. Ties in $z$ are broken by the order of the rows; if the rows may have been sorted by the response, they should first be permuted at random.

\subsection{The least squares test and its exact null distribution}\label{sec:kah3}

Let $s_j^2$ be the residual mean square of the least squares fit in part $P_j$, on $\nu=m-p-1$ degrees of freedom. The statistic is
\begin{equation}\label{eq:T1}
T_1=\frac{\max_j s_j^2}{\min_j s_j^2},
\end{equation}
and large values indicate heteroscedasticity. Unlike the Goldfeld--Quandt test, $T_1$ keeps the middle part and does not require the direction of the variance change to be known.

\begin{theorem}\label{thm:exact}
Under $H_0$ with normal errors, if the sorting variable $z$ is a function of the design or of variables independent of the errors (with ties broken independently of $y$), and if the design matrix of each part has full column rank $p+1$, $T_1$ follows Hartley's maximum $F$ distribution, denoted $\Fmax(k,\nu)$, with $k=3$ groups and $\nu=m-p-1$ degrees of freedom, for every $n$. Its distribution function is
\begin{equation}\label{eq:cdf}
P(T_1\le c)=k\int_0^1\bigl[G_\nu\{c\,G_\nu^{-1}(u)\}-u\bigr]^{k-1}\,du,\qquad c\ge1,\ k=3,
\end{equation}
where $G_\nu$ is the $\chi^2_\nu$ distribution function.
\end{theorem}

The proof is in Appendix~\ref{app:proofs}. The scale $\sigma^2$ cancels, the integrand in \eqref{eq:cdf} is bounded for every $\nu>0$, and critical values and $p$-values follow from one numerical integral. For $k=2$, \eqref{eq:cdf} reduces to $P(1/c\le F_{\nu,\nu}\le c)$, the two-sided variance-ratio test. The distribution of the maximum $F$ ratio was tabulated by \citet{Hartley1950} and \citet{David1952} for testing equality of variances across independent samples; \citet{ChuSutradhar1995} studied it under autocorrelation. In our setting the three samples are the three parts of an ordered regression, and their independence follows because the parts are disjoint and are chosen without reference to the errors. Ordering by $y$, by fitted values or by residuals is not covered.
\subsection{The robust test}\label{sec:robust}

The \KaHR{} test replaces each residual mean square by the raw LTS scale of the part. The LTS estimator \citep{Rousseeuw1984} minimizes the sum of the $h$ smallest squared residuals, where the coverage $\alpha\approx h/m$ is between 0.5 and 1. Its raw scale $\hat\sigma_j$ is the square root of the mean of the $h$ smallest squared residuals, multiplied by a consistency factor \citep{RousseeuwLeroy1987,PisonVanAelstWillems2002}. The statistic is
\begin{equation}\label{eq:T2}
T_2=\frac{\max_j\hat\sigma_j}{\min_j\hat\sigma_j},
\end{equation}
a ratio of scales, not of variances. Because the parts are disjoint and chosen independently of the errors, the three scales are independent under $H_0$ for any error distribution; because the consistency factor is common to the parts, it cancels, and $T_2$ does not depend on $\sigma$. Up to $m-h$ gross errors in a part leave its scale bounded (Section~\ref{sec:breakdown}).

The LTS scale is not an exact scaled $\chi^2$ variable, but it behaves like one with fewer degrees of freedom, because the trimming discards information. If $\hat\sigma_j^2/\sigma^2$ is approximated by $c_\alpha\chi^2_{\nu^*}/\nu^*$ with a constant $c_\alpha$, then
\begin{equation}\label{eq:approx}
T_2^2\ \dot\sim\ \Fmax(3,\nu^*).
\end{equation}
We choose $\nu^*$ by matching the variance of the logarithm, $\operatorname{var}\{\log(\chi^2_{\nu^*}/\nu^*)\}=\psi'(\nu^*/2)$, where $\psi'$ is the trigamma function; the constant $c_\alpha$ does not affect the logarithm's variance. Replacing $\nu$ by an effective number of degrees of freedom that reflects the extra variability of a scale estimator goes back to \citet{Box1953}, who corrected Bartlett's test for kurtosis in this way; see also \citet{Layard1973} and \citet{Shoemaker2003}. What is new here is the value of $\nu^*$ for the trimmed LTS scale.

\begin{proposition}\label{prop:nustar}
Let the errors be independent $N(0,\sigma^2)$ and $p$ fixed, and suppose that, as $m\to\infty$, the squared raw LTS scale of a part is asymptotically equivalent, up to a constant factor, to the lower-trimmed mean of the squared errors,
\[
\tilde\sigma^2=\frac1h\sum_{i=1}^{h}\varepsilon^2_{(i)},\qquad \varepsilon^2_{(1)}\le\dots\le\varepsilon^2_{(m)} .
\]
Let $Y=\varepsilon^2/\sigma^2\sim\chi^2_1$, $\xi=G_1^{-1}(\alpha)$ and $\tau=E(Y\mid Y\le\xi)$. Then $\nu^*/\nu\to r(\alpha)$, where
\begin{equation}\label{eq:ralpha}
r(\alpha)=\frac{2\,\alpha^2\tau^2}{\operatorname{var}\{(\xi-Y)_+\}}.
\end{equation}
\end{proposition}

Here $\nu^*/\nu$ is understood as the ratio of the asymptotic variances of $\log s_j^2$ and $\log\hat\sigma_j^2$; it coincides with the finite-sample definition above whenever $m\operatorname{var}(\log\hat\sigma_j^2)$ converges. The derivation, based on the influence function \citep{Hampel1974} of the trimmed mean \citep{Stigler1973}, is in Appendix~\ref{app:proofs}. The equivalence assumed in Proposition~\ref{prop:nustar} is the usual asymptotic description of the LTS scale \citep{RousseeuwLeroy1987,PisonVanAelstWillems2002,RousseeuwVanDriessen2006}. For regression it follows from the root-$m$ consistency of LTS: under symmetric errors the trimmed objective has zero gradient in the coefficients at the true value, so estimating them changes the trimmed mean by $O_p(1/m)$. The remaining steps are standard asymptotic arguments. As $\alpha\to1$, $r(\alpha)\to1$ and the least squares case is recovered. Numerically, $r(0.90)=0.781$, $r(0.75)=0.563$ and $r(0.50)=0.307$.

Because trimming inflates the asymptotic variance of $\log\hat\sigma_j^2$ by the factor $1/r(\alpha)$, $r(\alpha)$ is also the asymptotic relative efficiency of \KaHR{} with respect to \KaHIII{} at the normal model, against level breakdown points of about 3, 8 and 17\% of $n$ for $\alpha=0.9$, 0.75 and 0.5 (Section~\ref{sec:breakdown}); the default $\alpha=0.75$ balances the two. We use the raw LTS scale rather than a reweighted, S- or MM-scale because its null behaviour reduces to that of a trimmed variance, which gives the closed form \eqref{eq:ralpha}, and because its breakdown count is exact. The derivation does not use normality until the law of $Y$ is fixed: for another symmetric error law, \eqref{eq:ralpha} holds with that law in place of $\chi^2_1$. At $\alpha=0.75$ it gives 0.49 for $t_5$ errors, 0.44 for $t_3$ errors and 0.50 for normal errors contaminated by 10\% of $N(0,7^2)$, against 0.56 at the normal model. The normal-based $\nu^*$ therefore overstates the effective degrees of freedom by about 13 to 29\% under heavy tails and contamination, which explains the mildly liberal sizes reported under heavy tails and randomly placed outliers (Sections~\ref{sec:size} to \ref{sec:heavy}).

\begin{figure}[!htbp]
\centering
\includegraphics[width=\linewidth]{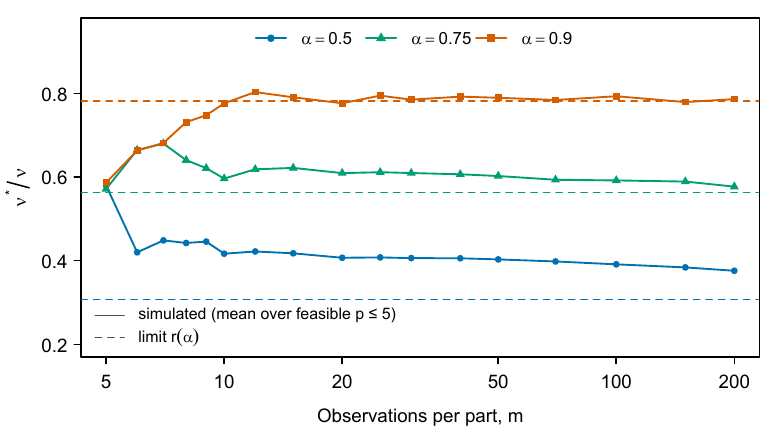}
\caption{Effective degrees of freedom of the raw LTS scale relative to least squares, $\nu^*/\nu$, against the part size $m$ for three coverages (solid lines, simulated, averaged over the feasible $p\le5$, all five for $m\ge15$). Dashed lines mark the limits $r(\alpha)$ of Proposition~\ref{prop:nustar}}\label{fig:nustar}
\end{figure}

The limit is reached slowly for strong trimming. We therefore estimated $\nu^*$ by simulation for $\alpha\in\{0.5,0.75,0.9\}$, $p=1,\dots,5$ and $m$ between 5 and 200, with 20{,}000 replications per cell. Fig.~\ref{fig:nustar} compares the simulated ratios with $r(\alpha)$; the values and the estimation procedure are given in Online Resource~1. For $\alpha=0.9$ they agree within 2\% from $m=20$; for $\alpha=0.75$ the simulated ratio exceeds $r(\alpha)$ by about 8 to 10\% for $m$ between 12 and 40, falling to 2.5\% at $m=200$; for $\alpha=0.5$ the gap is still 22\% at $m=200$. In practice we use the tabulated values, interpolated on $\log m$, and, beyond $m=200$, the mean simulated ratio over $m\ge50$; for $\alpha=0.5$ this extrapolated ratio lies above $r(0.5)$, so the Monte Carlo reference of the software is advisable there, as it is for parts of fewer than ten observations.

\subsection{Breakdown}\label{sec:breakdown}

A test breaks down when contamination can force its decision regardless of the data \citep{HeSimpsonPortnoy1990}. The statistic $T_2$ can be made arbitrarily large, and hence forced to reject, in two ways. One scale can be made to explode, which needs at least $m-h+1$ gross errors in one part. Or one scale can be made to implode, which needs $h-p-1$ further observations of one part placed on the hyperplane through $p+1$ of its points, for data in general position \citep{RousseeuwLeroy1987}. The smallest number of observations that can force rejection is therefore $b=\min(m-h+1,\,h-p-1)$, all in one part, with $h$ the coverage used by \pkg{robustbase}, and the level breakdown point is $b/n\approx(1-\alpha)/3$ for $\alpha\ge1/2$. When the contamination is spread evenly over the three parts, each part tolerates $b-1$ gross errors and the test tolerates about $(1-\alpha)n$ of them. For comparison, $T_1$, the Goldfeld--Quandt test and the score tests can be forced to reject by a single observation. For $\alpha=0.75$ and $m=20$, $h=15$ and $b=6$: a single part tolerates five gross errors. Ties or repeated regressor values within a part can lower the implosion count. At the default $\alpha=0.75$ the level breakdown point, about 8\% of $n$, is below the 10\% contamination of our simulations; the test does not break down there, and its level rises only mildly (7 to 11\%, Sections~\ref{sec:size} and \ref{sec:factorial}), because randomly placed outliers spread over the parts.

Conversely, contamination can hide heteroscedasticity only by making the part scales nearly equal. With fewer than $b$ replaced observations in a part, its LTS scale stays between bounds set by the good observations of that part, so when the variance differs strongly enough between the extreme parts, acceptance cannot be forced by fewer than $b$ observations either: the power breakdown point is at least about $(1-\alpha)/3$. For $T_1$ and the Goldfeld--Quandt test one suitably placed observation in the low-variance part suffices. Breakdown bounds say when a decision can be forced; Section~\ref{sec:masking} shows that concentrated outliers distort the test well before these bounds are reached.

The influence of a single observation is bounded as well. With the sorting variable fixed, an observation enters one part only and moves one scale only. Under symmetric errors the term due to estimating the coefficients vanishes, because the trimmed objective has zero gradient at the true coefficients. The influence function of the raw LTS scale then equals that of the trimmed variance, which is bounded: an error beyond the trimming point $\xi$ counts as if it were at $\xi$ (Appendix~\ref{app:proofs}; \citealp{Stigler1973}). Since $\log T_2$ is the range of the three log scales, a Lipschitz function of them, its influence in any contaminating direction is bounded by twice that of a log scale, and the level influence function of the test in the sense of \citet{HeritierRonchetti1994} is bounded at the symmetric error model. For $T_1$ the corresponding influence grows with the square of the error and has no bound. We report breakdown as the main robustness measure because it answers the practical question directly: how many gross errors can force a false rejection or hide a real one.

\subsection{A bootstrap reference}\label{sec:bootstrap}

The approximation \eqref{eq:approx} assumes normal errors. When the errors may be heavy-tailed, a residual bootstrap approximates the null distribution from the pooled LTS residuals. Let $e_i$ be the LTS residual of observation $i$ in its part and $j(i)$ the part that contains it. The standardized residuals $e_i/\hat\sigma_{j(i)}$ of the three LTS fits are pooled. Bootstrap responses $y^\star_i$ are drawn from them with replacement and assigned to the design, $T_2^\star$ is computed on each bootstrap sample, and the $p$-value is $(1+\#\{T_2^\star\ge T_2\})/(B+1)$. Because LTS is regression and scale equivariant, $T_2$ is invariant to the regression coefficients and to the scale, and the bootstrap responses need no fitted mean. The resampled residuals are not a sample from the error law (the retained residuals are shrunk by the fit and the trimmed ones are the largest), so the bootstrap is an approximation without a validity proof; it was conservative in every setting studied (Section~\ref{sec:size}). Because the trimmed residuals, which include any outliers, are resampled into all three parts, the bootstrap null distribution is more dispersed than that of a typical sample, which explains the conservativeness. The software uses $B=499$ by default; $B=199$ was used in the simulations to save computing time. Bootstrap references for heteroscedasticity tests were studied by \citet{GodfreyOrme1999}.

%======================================================================
\section{Simulation study}\label{sec:sim}

\subsection{Benchmark design}\label{sec:design}

The benchmark design follows the setting of most published comparisons: a variance that increases along one regressor. The regressors were drawn as independent $U(0,1)$ variables, standardized, and the observations were sorted by the first regressor $x_1$; all coefficients were one. We used $(n,p)\in\{(30,1),(60,1),(150,1),(90,2)\}$ and five scenarios. Under $H_0$ the errors were standard normal; $t_5$ scaled to unit variance; or standard normal with 10\% of the observations, at random positions, replaced by $N(0,7^2)$ errors. Under $H_1$ the error standard deviation was $\exp(0.35\,x_1)$, which varies by a factor of about 3.3 over the range of $x_1$, without or with the same 10\% contamination. Each cell had 1000 replications, so the Monte Carlo standard error of a rejection rate near 5\% is about 0.7 percentage points. All tests were applied to the same data sets, and the $p$-value of every test in every replication is stored with the code.

The competitors were the Goldfeld--Quandt test with the middle third omitted \citep{GoldfeldQuandt1965}; the Breusch--Pagan test in Koenker's studentized form \citep{BreuschPagan1979,Koenker1981}; White's test \citep{White1980}; the BAMSET test with three groups \citep{Ramsey1969}; the tests of \citet{EvansKing1988} and \citet{HarrisonMcCabe1979}; the modified Goldfeld--Quandt test (MGQ) of \citet{RanaMidiImon2008}; the robust Goldfeld--Quandt test (RGQ) of \citet{AlihOng2015}; White's test after an LTS outlier screen, following \citet{BerenguerRicoWilms2021}; the quantile-regression test of \citet{WilcoxKeselman2006}; the test of \citet{ZhouSongThompson2015}; and, as a check on the statistic rather than the partition, the Brown--Forsythe test \citep{BrownForsythe1974} on the residuals of the same three parts. Implementation details, including two corrections to published software, are given in Online Resource~1. The \KaHR{} test used $\alpha=0.75$, the default of the package, and the reference \eqref{eq:approx}; its bootstrap version used $B=199$.

\subsection{Null distributions}\label{sec:nullfit}

\begin{figure}[!htbp]
\centering
\includegraphics[width=\linewidth]{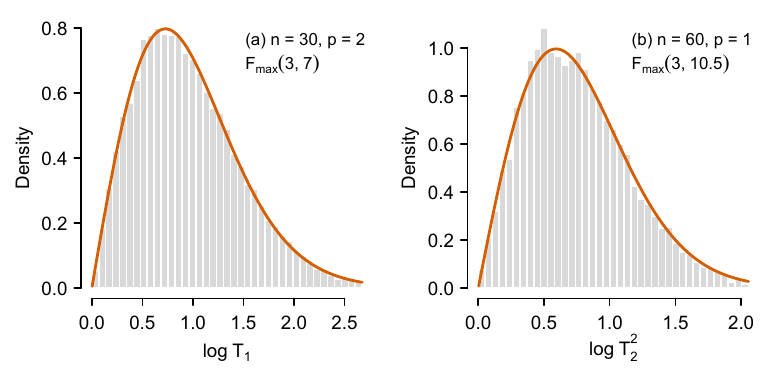}
\caption{Simulated null distributions (grey histograms) and references (lines): (a) $\log T_1$ with the exact $\Fmax(3,7)$ density, $n=30$, $p=2$, 20{,}000 replications; (b) $\log T_2^2$ at the default $\alpha=0.75$ with the $\Fmax(3,\nu^*)$ approximation, $n=60$, $p=1$, 10{,}000 replications}\label{fig:null}
\end{figure}

Fig.~\ref{fig:null} compares simulated null distributions with the references. For $T_1$, a further simulation with $10^5$ replications at five settings, including $n=31$, gave Kolmogorov--Smirnov $p$-values between 0.54 and 0.92. The tabulated 5\% and 1\% points \citep{Hartley1950,David1952} for $k=3$ and $\nu=4$, 15.5 and 37, were reproduced as 15.46 and 36.70. For $T_2$ at the default $\alpha=0.75$, we ran twelve settings with $n$ between 15 and 300 and $p$ between 1 and 3, with 10{,}000 replications each, using the LTS fits and the values of $\nu^*$ of the package. The true sizes of the nominal 5\% test from \eqref{eq:approx} were between 4.4\% and 5.6\%, and those of the nominal 1\% test between 0.6\% and 1.2\% (Monte Carlo standard errors 0.2 and 0.1 points). The Kolmogorov--Smirnov test rejected the fit of the whole distribution only for the two smallest part sizes, $m=5$ and $m=8$ ($p=0.016$ and $0.007$); in the other ten settings its $p$-values were between 0.13 and 0.97.

\subsection{Size}\label{sec:size}

\begin{figure}[!htbp]
\centering
\includegraphics[width=\linewidth]{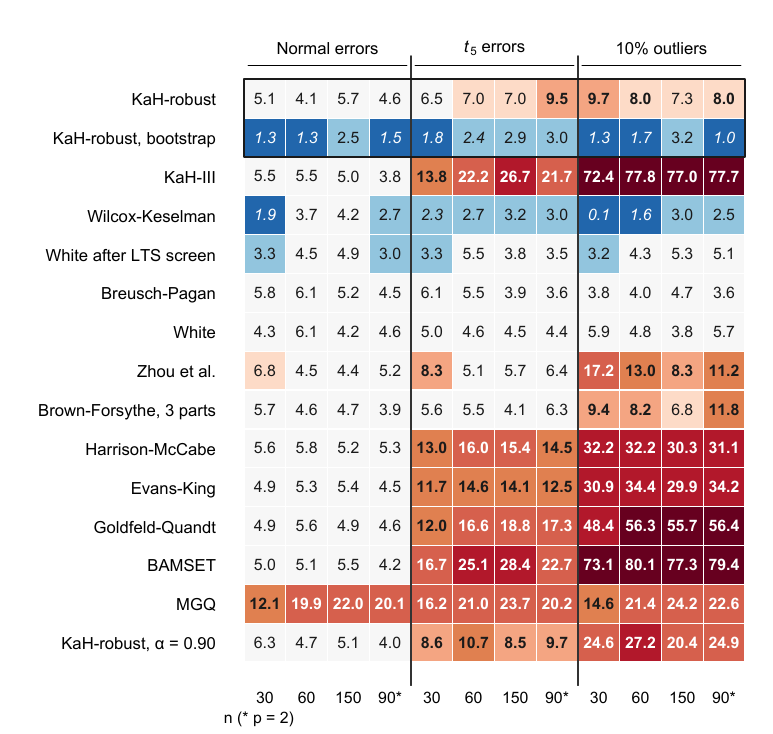}
\caption{Empirical size (\%) at the nominal 5\% level under $H_0$, 1000 replications per cell (Monte Carlo standard error about 0.7 percentage points). Columns give $n$ ($p=1$, except $p=2$ where marked); values above 7.5\% are in bold and values below 2.5\% in italics. The framed rows are \KaHR{} with the default reference and with the bootstrap reference}\label{fig:size}
\end{figure}

Fig.~\ref{fig:size} shows the empirical sizes of all tests. Under normal errors, the \KaHIII{}, \KaHR{}, Goldfeld--Quandt, Breusch--Pagan, White, BAMSET, Evans--King and Harrison--McCabe tests held the nominal level. The MGQ test did not: its size was between 12 and 22\%, because the ratio of median squared residuals is not $F$-distributed. Under contamination, the tests that compare least squares variances across ordered groups rejected far too often: \KaHIII{} 72 to 78\%, Goldfeld--Quandt 48 to 56\%, BAMSET 73 to 80\%, Evans--King and Harrison--McCabe around 30\%. The score tests of Breusch--Pagan and White kept their level, because randomly placed outliers do not change the regression of squared residuals on the regressors. Under contamination \KaHR{} rejected 7 to 10\% of true null hypotheses, which is mildly liberal. The Brown--Forsythe test on the same three parts, which uses absolute deviations from the median, rejected 7 to 12\%, but it had little power under contamination (Section~\ref{sec:power}). With $\alpha=0.90$ the trimming (10\%) is no larger than the contamination rate, and \KaHR{} rejected 20 to 27\%; the default $\alpha=0.75$ leaves a margin. With $\alpha=0.5$ the sizes were 4.5 to 5.3\% under normal errors, 6.5 to 7.5\% with outliers and 6.0 to 7.2\% under $t_5$ errors. The bootstrap version of \KaHR{} and the Wilcox--Keselman test were conservative.

\subsection{A factorial comparison}\label{sec:factorial}

\begin{figure}[!htbp]
\centering
\includegraphics[width=\linewidth]{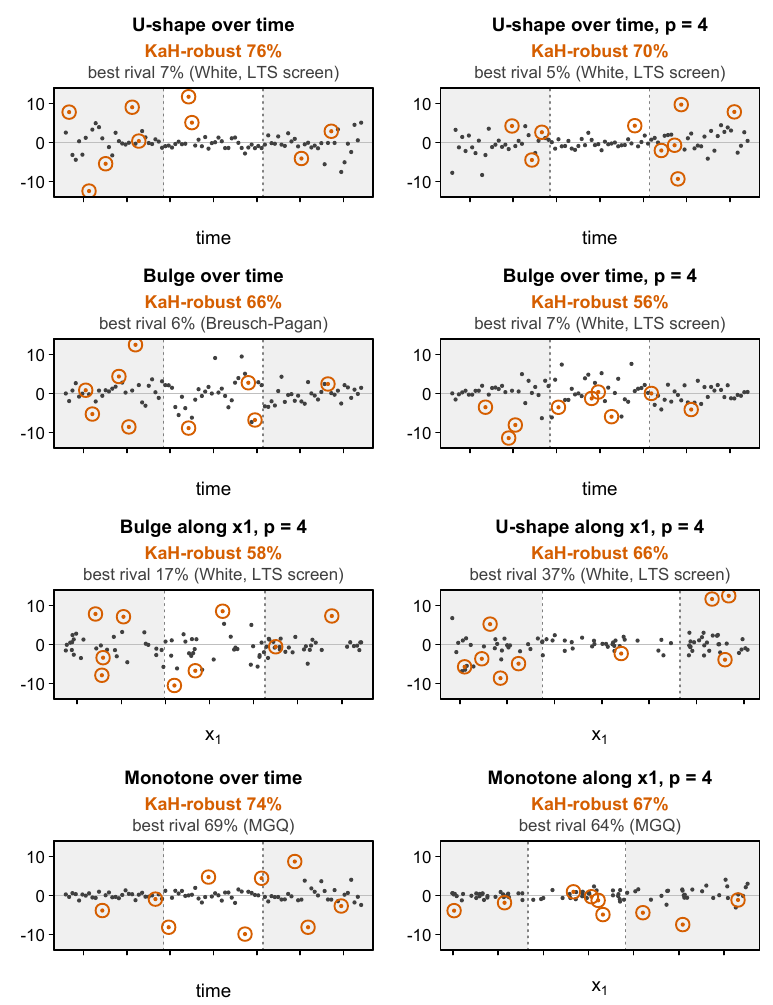}
\caption{The eight settings of the factorial study ($n=90$) in which \KaHR{} had the largest margin over every competing test. Each panel shows one example data set: the errors against the sorting variable, with the three parts shaded and the 10\% outliers circled. The printed values are size-adjusted powers from 1000 replications of that setting: \KaHR{} and the best of the nine other tests of the factorial study (eight when $p=4$); differences under about 4 points are within Monte Carlo error. In the time-ordered panels the regressor-based tests were not given the time variable}\label{fig:scenarios}
\end{figure}

To compare the tests beyond a single setting, we ran a full factorial design with five factors:
\begin{itemize}
  \item the shape of the error standard deviation as a function of the ordering variable $z$: monotone, $\exp(0.35z)$; U-shaped, $1+z^2$, high at both ends and low in the middle; or bulge-shaped, $1+3\exp(-z^2)$, low at both ends and high in the middle;
  \item the ordering variable: the first regressor, or the time order of the observations, independent of the regressors;
  \item the number of regressors, $p=1$ or 4;
  \item contamination: none, or 10\% of $N(0,7^2)$ vertical outliers at random positions;
  \item the sample size, $n\in\{45,90,150\}$.
\end{itemize}
This gives 72 heteroscedastic cells and 24 matching null cells, each with 1000 replications. The Harrison--McCabe and Zhou et al.\ tests and the bootstrap version of \KaHR{} were left out because they were dominated or conservative in Section~\ref{sec:size}, the Brown--Forsythe test on the parts because it lost its power under contamination, and the Wilcox--Keselman test, whose quantile-regression bootstrap is slow, was run for $p=1$ only. Tests that sort the data (the KaH tests, Goldfeld--Quandt, BAMSET and MGQ) used the same ordering variable. The Breusch--Pagan, White, White-after-LTS-screen, Evans--King (with $x_1$ as deflator) and Wilcox--Keselman tests used the regressors, as they are defined, and were not given the ordering variable; in the time-ordered settings this handicaps them, and the two orderings are reported separately below. We report size-adjusted power: each test's rejection threshold is taken from its own null distribution in the matching null cell (for \KaHIII{}, whose $p$-values underflow to zero in some contaminated samples, from the statistic itself), so differences in size do not affect the comparison. With 1000 replications the Monte Carlo standard error of a size-adjusted power near 70\% is about 1.5 points, and that of a difference between two tests about 2 points; differences under 4 points are within Monte Carlo error. In the 24 null cells \KaHR{} rejected 4 to 6\% of true null hypotheses on clean data and 7 to 11\% with outliers; \KaHIII{}, BAMSET, Goldfeld--Quandt and Evans--King rejected 30 to 81\% with outliers, and MGQ 12 to 24\% throughout. Fig.~\ref{fig:scenarios} shows example data sets for the eight settings in which \KaHR{} had the largest margin over every competing test.

\begin{figure}[!htbp]
\centering
\includegraphics[width=\linewidth]{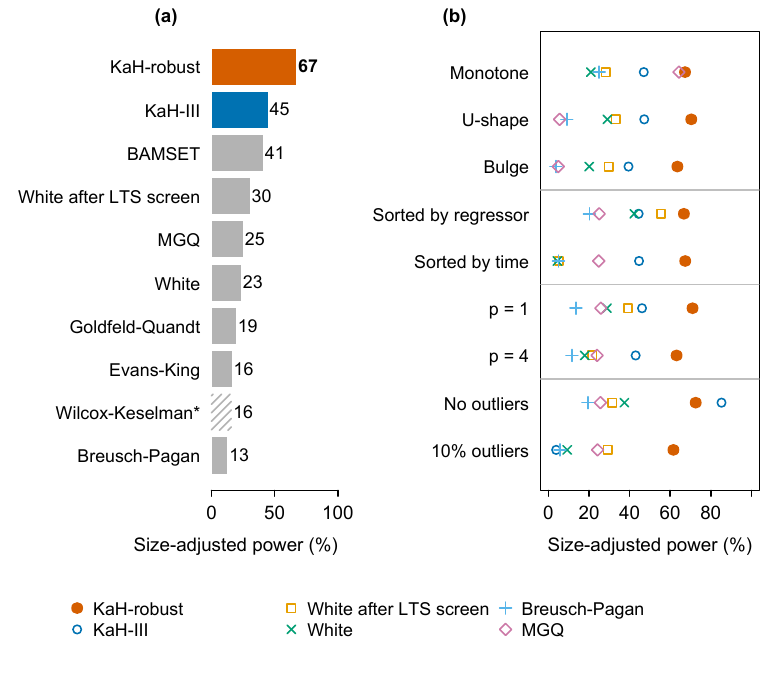}
\caption{Factorial comparison over 72 heteroscedastic settings: three variance shapes, two ordering variables, $p\in\{1,4\}$, with and without 10\% outliers, $n\in\{45,90,150\}$; 1000 replications per setting. (a) Mean size-adjusted power of each test over all settings; the hatched Wilcox--Keselman bar covers the $p=1$ settings only, over which \KaHR{} averaged 71\% and White's test after an LTS screen 39\%. (b) Mean size-adjusted power at each level of four of the five factors (sample size is given in the text) for six tests}\label{fig:factorial}
\end{figure}

\KaHR{} had the highest mean size-adjusted power over the 72 cells, 67\%, against 45\% for \KaHIII{}, 41\% for BAMSET and at most 30\% for every other test (Fig.~\ref{fig:factorial}a). It won more cells than any other test and had the highest mean power at every level of every factor except two: without outliers, where \KaHIII{} (85\%) and BAMSET (79\%) led, and at $n=45$, where the two KaH tests tied at 33\% (at $n=90$ and 150, \KaHR{} reached 74 and 94\%). Fig.~\ref{fig:factorial}b shows where the advantage comes from. For U- and bulge-shaped variance, the Goldfeld--Quandt, Breusch--Pagan, Evans--King, Wilcox--Keselman and MGQ tests had mean power below 15\%, because each looks for a variance that moves in one direction, whereas \KaHR{} reached 70\% and 64\%. When the variance changed along time, every test built on the regressors stayed near 5\%, while \KaHR{}, sorted by time, reached 67\%; when it changed along a regressor, the best of them, White's test after an LTS screen, reached 56\%, against 67\% for \KaHR{}. When the ordering variable is one of its regressors, White's test can represent U- and bulge-shaped patterns, as it does in the settings ordered by $x_1$ (53 and 36\%, and 61 and 54\% after an LTS screen), so the time-ordered contrast reflects the information the tests are given. With four regressors, \KaHR{} kept 63\% while White's test fell to 18\%, because its auxiliary regression grows to 14 terms. With outliers and the variance changing along $x_1$, where all tests were given the variable, \KaHR{} kept 62\% against 53\% for the best competitor, White's test after an LTS screen.

The two versions complement each other. At $n=90$, \KaHR{} had the highest size-adjusted power of all tests in 9 of the 24 settings, all of them with outliers. \KaHIII{} had it in 6 of the 12 settings without outliers and tied with White's test in a seventh. One of the two proposed tests was thus the most powerful in 15 of the 24 settings, and tied in one more. This suggests a simple rule for practice: \KaHIII{} when the errors are close to normal and free of outliers, and \KaHR{} otherwise. The contamination status is rarely known, and applying \KaHIII{} to contaminated or heavy-tailed data destroys its level (Fig.~\ref{fig:size}, Table~\ref{tab:heavy}); when in doubt, \KaHR{} is the default.

\begin{figure}[!htbp]
\centering
\includegraphics[width=\linewidth]{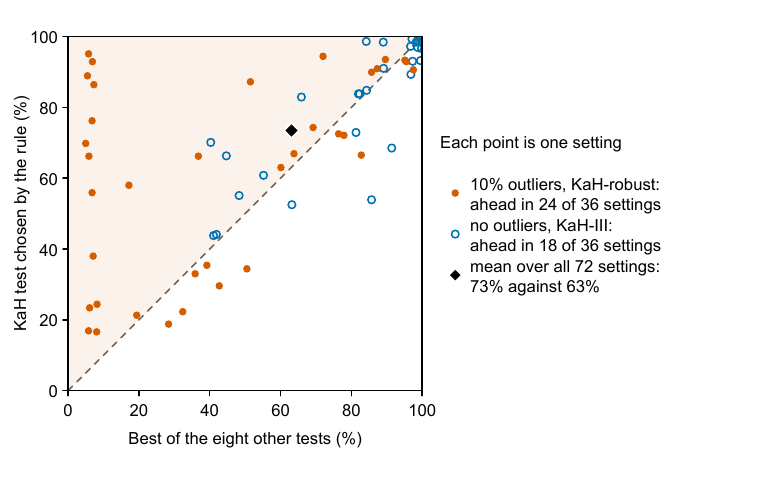}
\caption{The usage rule against the best other test, setting by setting. Each point is one of the 72 heteroscedastic settings of the factorial study; the vertical axis gives the size-adjusted power of the KaH test chosen by the rule (\KaHIII{} without outliers, \KaHR{} with 10\% outliers; the rule uses the true contamination status), and the horizontal axis that of the most powerful of the eight other tests in the same setting (seven when $p=4$). Points above the diagonal are settings in which the KaH test is more powerful than all of them; differences under 4 points are within Monte Carlo error. The black diamond marks the means over all settings}\label{fig:rule}
\end{figure}

Fig.~\ref{fig:rule} tests this rule against the strongest possible opponent: in each setting, the best of the eight other tests, a test no analyst could choose in advance (the rule, in turn, assumes that the presence of outliers is known). The rule was ahead in 42 of the 72 settings (24 with outliers and 18 without), by 4 points or more in 26 of them, and behind by 4 points or more in 14; its mean size-adjusted power was 73\% against 63\% for this setting-by-setting best and 41\% for the best single competing test, BAMSET. \KaHR{} alone, which needs no knowledge of the outliers, averaged 67\%, also above the setting-by-setting best, and was ahead of it in 25 settings, by 4 points or more in 19.

\subsection{Heavy-tailed errors}\label{sec:heavy}

Heavy tails produce outliers without any being planted. We therefore repeated the factorial design without contamination but with $t_3$ errors scaled to unit variance, a distribution whose fourth moment is infinite (Table~\ref{tab:heavy}). The least squares ordered-group tests lost their level. At the nominal 5\% level, \KaHIII{} and BAMSET rejected 28 to 53\% of true null hypotheses, Goldfeld--Quandt 22 to 39\% and Evans--King 17 to 24\%, as expected from the sensitivity of variance tests to the kurtosis of the errors \citep{Box1953}. \KaHR{} rejected 8 to 11\% (Monte Carlo standard error about 0.9 points), and its bootstrap version 0.5 to 3.8\%.

With the levels equalized, \KaHR{} had a mean size-adjusted power of 64\%, against 35\% for \KaHIII{} and at most 24\% for every other test. It was the most powerful test in 28 of the 36 settings, by 4 points or more in 23, and led on average for all three variance shapes. In the settings with a variance monotone in $x_1$ alone, the Evans--King (66\%) and Wilcox--Keselman (67\%) tests were ahead of \KaHR{} (62\%), and MGQ came close (60\%) but rejected 15 to 26\% of true null hypotheses at the nominal level. The advantage of \KaHR{} found with planted outliers in Section~\ref{sec:factorial} therefore holds, at a similar margin, when the outliers come from the error distribution itself.

\begin{table}[!htbp]
\centering
\caption{Heavy-tailed errors: $t_3$ errors scaled to unit variance and no planted outliers, with the factorial design of Section~\ref{sec:factorial} otherwise unchanged (two ordering variables, $p\in\{1,4\}$, $n\in\{45,90,150\}$); 1000 replications per cell. Size: rejection rate (\%) at the nominal 5\% level over the 12 null cells. Power: mean size-adjusted power (\%) over the 36 heteroscedastic cells, overall and by variance shape. Wins: cells in which the test had the highest size-adjusted power. Rows are sorted by mean size-adjusted power; the Wilcox--Keselman test, whose quantile-regression bootstrap is slow, was run for $p=1$ only. The bootstrap reference of \KaHR{} ($B=199$) had size 0.5 to 3.8\% (mean 2.5\%) in the same null cells}\label{tab:heavy}
\small\setlength{\tabcolsep}{4pt}
\begin{tabular}{@{}lrrrrrrr@{}}
\toprule
 & \multicolumn{2}{c}{Size} & \multicolumn{4}{c}{Size-adjusted power} & \\
\cmidrule(lr){2-3}\cmidrule(lr){4-7}
Test & range & mean & all & monotone & U & bulge & Wins \\
\midrule
KaH-robust & 8--11 & 9 & 64 & 63 & 68 & 61 & 28 \\
KaH-III & 28--53 & 42 & 35 & 34 & 45 & 26 & 1 \\
BAMSET & 28--52 & 42 & 24 & 23 & 27 & 23 & 0 \\
MGQ & 15--26 & 22 & 24 & 61 & 6 & 5 & 1 \\
White, LTS screen & 2--5 & 4 & 22 & 19 & 25 & 22 & 1 \\
Goldfeld--Quandt & 22--39 & 30 & 19 & 46 & 6 & 6 & 0 \\
White & 4--10 & 7 & 19 & 17 & 27 & 12 & 1 \\
Evans--King & 17--24 & 20 & 17 & 36 & 7 & 7 & 3 \\
Wilcox--Keselman & 2--4 & 3 & 15 & 36 & 4 & 5 & 1 \\
Breusch--Pagan & 3--5 & 4 & 12 & 24 & 9 & 3 & 0 \\
\bottomrule
\end{tabular}
\end{table}

\subsection{Concentrated and asymmetric contamination}\label{sec:masking}

The factorial study places outliers at random positions. Section~\ref{sec:breakdown} shows that the weak point of \KaHR{} is contamination concentrated in one part, which is also the masking case of the Introduction. Table~\ref{tab:masking} reports three further designs with a variance monotone in $x_1$: outliers only in the low-variance part, in 10 to 30\% of it; outliers at random positions at rates from 5 to 30\%; and one-sided $N(10,1)$ outliers. \KaHIII{} lost its level in every contaminated design. With outliers in 10\% of the low-variance part, \KaHR{} was mildly liberal (7 to 10\%), as with random outliers, and kept much of its power. With 20 and 30\% of the part, its size rose to 15 to 76\% and its size-adjusted power fell to at most 26\%, already at 20\%, which is within the breakdown count of Section~\ref{sec:breakdown}; $\alpha=0.5$ reduced the size distortion (9 to 40\%) but not the loss of power. At rates of 20 to 30\% at random positions \KaHR{} was liberal (13 to 45\%), and with one-sided outliers mildly so (9 to 12\%). White's test after an LTS screen of the whole sample, which removes outliers before testing, held its level (4 to 8\%) in every design of Table~\ref{tab:masking} and had clearly more size-adjusted power when the outliers were concentrated in the low-variance part, exceeded 10\% of $n$ or were one-sided; with 5 to 10\% of outliers at random positions the two tests were close. In a separate study (not in the table), a contiguous block of outliers shifted by four error standard deviations, 10\% of $n$ placed in the middle part (30\% of that part), gave sizes of 46 to 99\% at $\alpha=0.75$ and 15 to 55\% at $\alpha=0.5$ for $n$ between 30 and 150. The trimming within each part thus protects \KaHR{} against outliers spread along the ordering, as in Sections~\ref{sec:size} to \ref{sec:heavy}, but when outliers may cluster in one region of the ordering or exceed about 10\% of the data, a global outlier screen is the safer choice.

\begin{table}[!htbp]
\centering
\caption{Concentrated and asymmetric contamination: rejection rate (\%) at the nominal 5\% level under $H_0$ and size-adjusted power (\%) under a variance monotone in $x_1$ (standard deviation $\exp(0.35x_1)$), $p=1$, 1000 replications per cell. Outliers are $N(0,7^2)$ errors, or $N(10,1)$ for the asymmetric case. R75 and R50: \KaHR{} with $\alpha=0.75$ and 0.5; White-LTS: White's test after an LTS screen. Results for $n=45$ are in the deposit}\label{tab:masking}
\scriptsize\setlength{\tabcolsep}{2.5pt}
\begin{tabular}{@{}lrrrrrrrrr@{}}
\toprule
 & & \multicolumn{4}{c}{Size, $H_0$} & \multicolumn{4}{c}{Power, size-adjusted} \\
\cmidrule(lr){3-6}\cmidrule(lr){7-10}
Contamination & $n$ & R75 & R50 & KaH-III & White-LTS & R75 & R50 & KaH-III & White-LTS \\
\midrule
\multicolumn{10}{@{}l}{\emph{Outliers in the low-variance part}} \\
0\% of the part & 90 & 4 & 5 & 6 & 4 & 85 & 62 & 96 & 79 \\
 & 150 & 5 & 6 & 5 & 4 & 97 & 83 & 100 & 96 \\
10\% of the part & 90 & 7 & 6 & 87 & 4 & 50 & 37 & 2 & 72 \\
 & 150 & 10 & 8 & 96 & 4 & 81 & 60 & 2 & 93 \\
20\% of the part & 90 & 21 & 12 & 99 & 5 & 14 & 13 & 1 & 61 \\
 & 150 & 30 & 18 & 100 & 6 & 26 & 25 & 3 & 86 \\
30\% of the part & 90 & 54 & 26 & 100 & 6 & 2 & 3 & 3 & 47 \\
 & 150 & 76 & 40 & 100 & 8 & 0 & 3 & 3 & 66 \\
\multicolumn{10}{@{}l}{\emph{Outliers at random positions}} \\
5\% of $n$ & 90 & 6 & 7 & 79 & 5 & 79 & 54 & 14 & 75 \\
 & 150 & 6 & 5 & 83 & 5 & 97 & 85 & 16 & 97 \\
10\% of $n$ & 90 & 7 & 5 & 77 & 4 & 73 & 59 & 10 & 75 \\
 & 150 & 8 & 7 & 72 & 5 & 92 & 79 & 8 & 95 \\
20\% of $n$ & 90 & 19 & 10 & 61 & 4 & 44 & 36 & 5 & 59 \\
 & 150 & 13 & 8 & 61 & 6 & 78 & 74 & 5 & 82 \\
30\% of $n$ & 90 & 35 & 10 & 46 & 5 & 23 & 40 & 5 & 36 \\
 & 150 & 35 & 13 & 47 & 6 & 34 & 59 & 5 & 53 \\
\multicolumn{10}{@{}l}{\emph{Asymmetric outliers}} \\
10\% of $n$, one-sided & 90 & 11 & 9 & 39 & 5 & 72 & 51 & 4 & 81 \\
 & 150 & 9 & 6 & 43 & 4 & 94 & 83 & 6 & 99 \\
\bottomrule
\end{tabular}
\end{table}

\subsection{Power under a monotone variance}\label{sec:power}

\begin{figure}[!htbp]
\centering
\includegraphics[width=\linewidth]{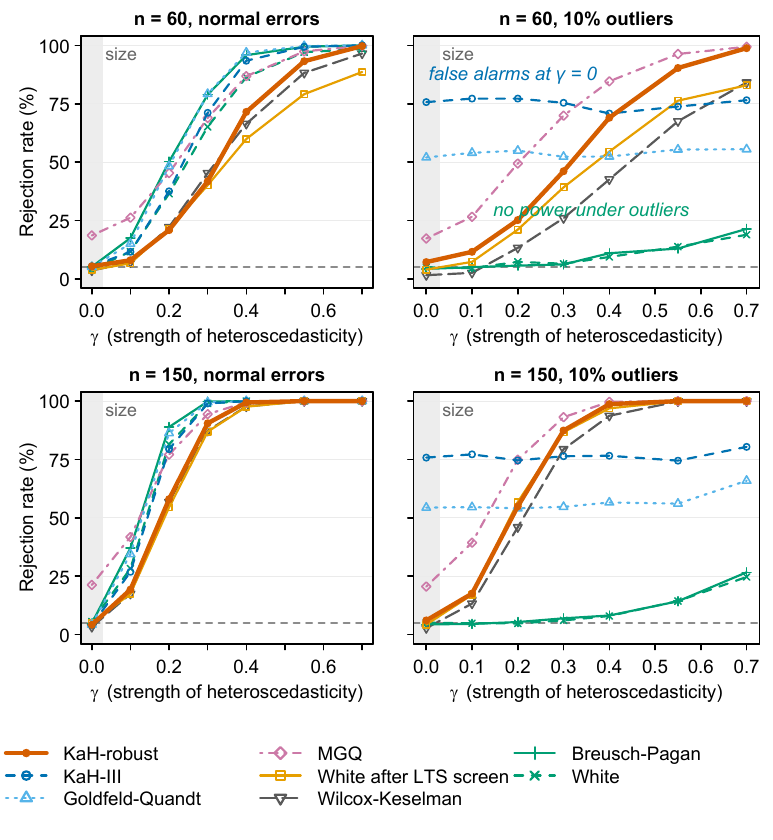}
\caption{Power curves: rejection rate at the nominal 5\% level against the strength $\gamma$ of heteroscedasticity, with error standard deviation $\exp(\gamma x_1)$, $p=1$, 1000 replications per point and the same data sets for every test. The shaded column $\gamma=0$ is the size; the dashed line marks 5\%. With 10\% outliers (right), curves that start far above 5\% belong to tests that reject true null hypotheses}\label{fig:power}
\end{figure}

The benchmark design of Section~\ref{sec:design} looks at the monotone case more closely. Table~\ref{tab:power} reports power at one strength of heteroscedasticity, and Fig.~\ref{fig:power} traces power curves as the strength grows. With normal errors, the tests designed for a variance that is monotone in $x_1$ were the most powerful: Evans--King, Harrison--McCabe, Breusch--Pagan and Goldfeld--Quandt. \KaHIII{}, which does not assume a direction, followed (at $(n,p)=(90,2)$ it passed the Breusch--Pagan test, 95.3 against 93.1\%), ahead of BAMSET and White's test. It was well behind at $n=30$, the gap narrowed as $n$ grew, and at $n=150$ all these tests had power near 100\%. \KaHR{} paid for its robustness, with power 28\% against 44\% for \KaHIII{} at $n=30$; the gap closed as $n$ grew. With contamination the ranking changed: the Breusch--Pagan and White tests lost almost all power (6 to 8\%), and among the tests whose size stayed at or below 10\%, \KaHR{} had the highest power at the nominal level in every configuration, partly because the others were conservative.

The power curves in Fig.~\ref{fig:power} show the same contrast over the whole range of heteroscedasticity. With 10\% outliers the curves of \KaHIII{} and the Goldfeld--Quandt test are flat at about 76 and 54\%, so their rejections carry no information about the variance, and those of the Breusch--Pagan and White tests stay below 27\%. Among the robust tests \KaHR{} rises fastest, from a slightly higher start (7\% against 2\% for Wilcox--Keselman), reaching 90\% at $\gamma=0.55$ and $n=60$ against 76\% for White's test after an LTS screen and 68\% for the Wilcox--Keselman test; the MGQ curve starts at 17 to 21\%.

\begin{table}[!htbp]
\centering
\caption{Empirical power (\%) at the nominal 5\% level; error standard deviation $\exp(0.35\,x_1)$; columns give $(n,p)$. Rates for tests whose size is not controlled in Fig.~\ref{fig:size} are not valid power}\label{tab:power}
\scriptsize
\setlength{\tabcolsep}{2pt}
\begin{tabular}{lrrrrrrrr}
\toprule
& \multicolumn{4}{c}{Normal errors} & \multicolumn{4}{c}{10\% outliers} \\
\cmidrule(lr){2-5}\cmidrule(lr){6-9}
Test & 30,1 & 60,1 & 150,1 & 90,2 & 30,1 & 60,1 & 150,1 & 90,2 \\
\midrule
KaH-III & 43.8 & 83.3 & 100.0 & 95.3 & 70.9 & 73.2 & 73.4 & 76.2 \\
KaH-robust, 0.75 & 27.7 & 57.4 & 98.3 & 77.6 & 30.6 & 55.4 & 95.9 & 75.4 \\
KaH-robust, bootstrap & 12.3 & 37.7 & 95.6 & 60.1 & 7.4 & 30.4 & 88.5 & 49.8 \\
KaH-robust, 0.90 & 36.2 & 73.0 & 99.8 & 89.2 & 48.0 & 69.7 & 95.5 & 81.0 \\
Goldfeld--Quandt & 54.1 & 89.9 & 100.0 & 97.6 & 51.5 & 55.0 & 52.5 & 55.3 \\
Breusch--Pagan & 54.2 & 90.3 & 100.0 & 93.1 & 8.0 & 6.4 & 6.1 & 6.1 \\
White & 39.5 & 77.3 & 99.9 & 79.8 & 7.7 & 7.5 & 6.2 & 7.5 \\
MGQ & 50.7 & 77.8 & 98.7 & 88.6 & 47.3 & 77.4 & 98.1 & 87.6 \\
White, LTS screen & 15.1 & 49.4 & 95.4 & 48.9 & 16.2 & 46.2 & 94.6 & 49.2 \\
Wilcox--Keselman & 26.2 & 56.1 & 96.4 & 74.6 & 7.0 & 36.7 & 91.3 & 58.9 \\
BAMSET & 40.0 & 81.6 & 100.0 & 90.9 & 74.3 & 76.9 & 74.0 & 77.8 \\
Evans--King & 74.4 & 96.8 & 100.0 & 99.7 & 46.1 & 43.9 & 44.3 & 48.3 \\
Harrison--McCabe & 65.8 & 92.3 & 99.9 & 97.8 & 42.6 & 40.0 & 40.2 & 42.1 \\
Zhou et al. & 15.4 & 20.0 & 27.3 & 13.5 & 17.0 & 14.0 & 9.9 & 12.9 \\
Brown--Forsythe, parts & 30.5 & 71.3 & 99.8 & 90.1 & 15.7 & 18.0 & 26.9 & 26.4 \\
\bottomrule
\end{tabular}
\end{table}

Table~\ref{tab:sizeadj} adds size-adjusted power for this monotone case. With the levels equalized, \KaHR{} did not lead in this monotone case: the Wilcox--Keselman test was ahead in all four configurations (by 0.2 to 7 points), and MGQ, White's test after an LTS screen and the bootstrap version of \KaHR{} were ahead in three of the four, mostly by less than 5 points. At the nominal level the Wilcox--Keselman test and the bootstrap version of \KaHR{} were conservative (sizes 0.1 to 3\% and 1 to 3\%) and lost that advantage, MGQ rejected 15 to 24\% of true null hypotheses, and White's test after an LTS screen held its level but had less power. The monotone case is the one in which the competitors are strongest; over the factorial design of Section~\ref{sec:factorial}, \KaHR{} led by a wide margin.

\begin{table}[!htbp]
\centering
\caption{Tests under 10\% outliers: size under $H_0$ and size-adjusted power under $H_1$ (rejection threshold set to the 5\% quantile of the test's own $p$-values under $H_0$ with outliers); power at the nominal level is in Table~\ref{tab:power}; columns give $(n,p)$; 1000 replications on the same data sets for every test}\label{tab:sizeadj}
\scriptsize
\setlength{\tabcolsep}{2pt}
\begin{tabular}{lrrrrrrrr}
\toprule
& \multicolumn{4}{c}{Size, $H_0$} & \multicolumn{4}{c}{Power, size-adjusted} \\
\cmidrule(lr){2-5}\cmidrule(lr){6-9}
Test & 30,1 & 60,1 & 150,1 & 90,2 & 30,1 & 60,1 & 150,1 & 90,2 \\
\midrule
KaH-robust & 9.7 & 8.0 & 7.3 & 8.0 & 19.8 & 45.1 & 94.2 & 67.8 \\
KaH-robust, bootstrap & 1.3 & 1.7 & 3.2 & 1.0 & 22.5 & 48.9 & 92.7 & 68.4 \\
White, LTS screen & 3.2 & 4.3 & 5.3 & 5.1 & 20.3 & 49.5 & 94.5 & 48.9 \\
Wilcox--Keselman & 0.1 & 1.6 & 3.0 & 2.5 & 26.6 & 50.9 & 94.4 & 69.7 \\
MGQ & 14.6 & 21.4 & 24.2 & 22.6 & 27.9 & 50.1 & 89.1 & 68.1 \\
Breusch--Pagan & 3.8 & 4.0 & 4.7 & 3.6 & 9.1 & 7.4 & 6.2 & 7.5 \\
White & 5.9 & 4.8 & 3.8 & 5.7 & 6.1 & 7.6 & 8.1 & 6.5 \\
Goldfeld--Quandt & 48.4 & 56.3 & 55.7 & 56.4 & 10.7 & 9.9 & 6.2 & 9.2 \\
BAMSET & 73.1 & 80.1 & 77.3 & 79.4 & 6.7 & 8.9 & 5.1 & 7.6 \\
\bottomrule
\end{tabular}
\end{table}

\subsection{The robust Goldfeld--Quandt test of Alih and Ong}\label{sec:rgq}

The RGQ test ranks the observations by a robust distance computed from the regressors and the response jointly. Implemented as described in its paper, it rejected 85 to 100\% of true null hypotheses, because ranking by a distance that involves the response places the observations with large residuals in one group by construction. Ranked by $x_1$ instead, it held its level under normal errors but not under contamination (12 to 30\%) or $t_5$ errors (9 to 14\%). The worked example of \citet{AlihOng2015} could not be reproduced from the published description, so our implementation is a good-faith reading of it; the details and the full results are in Online Resource~1.

%======================================================================
\section{Examples}\label{sec:data}

We applied the tests to four benchmark data sets used by the authors of the competing robust tests, each without and with the outliers those authors planted:
\begin{itemize}
  \item the housing expenditure data of \citet{PindyckRubinfeld1998} with the two outliers of \citet{RanaMidiImon2008} ($n=20$);
  \item the savings and income data of \citet{Koutsoyiannis2001} with the four outliers of \citet{AlihOng2015} ($n=31$);
  \item the restaurant sales data of \citet{MontgomeryPeckVining2001} with the three outliers of \citet{AlihOng2015} ($n=30$);
  \item the consumption expenditure data of \citet{Gujarati2003} with the three outliers of \citet{AlihOng2015} ($n=30$).
\end{itemize} In each case the observations were sorted by the single regressor. Table~\ref{tab:real} gives the $p$-values.

\begin{table}[!htbp]
\centering
\caption{$p$-values on four benchmark data sets, without and with the outliers planted by the authors of the competing robust tests ($n=20$, 31, 30 and 30). The bootstrap reference used $B=999$. A dash marks a test that could not be computed: in the housing data with outliers a matrix in the outlier screen of RGQ is singular, because the regressor takes few distinct values}\label{tab:real}
\scriptsize
\setlength{\tabcolsep}{2pt}
\begin{tabular}{lcccccccc}
\toprule
& \multicolumn{2}{c}{Housing} & \multicolumn{2}{c}{Savings} & \multicolumn{2}{c}{Restaurant} & \multicolumn{2}{c}{Consumption} \\
\cmidrule(lr){2-3}\cmidrule(lr){4-5}\cmidrule(lr){6-7}\cmidrule(lr){8-9}
Test & clean & outliers & clean & outliers & clean & outliers & clean & outliers \\
\midrule
KaH-III & 0.013 & $<$0.001 & 0.030 & 0.002 & 0.088 & $<$0.001 & 0.143 & 0.005 \\
KaH-robust & 0.011 & 0.009 & 0.032 & 0.014 & 0.024 & 0.004 & 0.186 & 0.291 \\
KaH-robust, bootstrap & 0.085 & 0.183 & 0.047 & 0.199 & 0.084 & 0.267 & 0.330 & 0.546 \\
Goldfeld--Quandt & 0.006 & 0.694 & 0.012 & 0.787 & 0.037 & 0.025 & 0.062 & 0.677 \\
Breusch--Pagan & 0.007 & 0.557 & 0.001 & 0.357 & 0.037 & 0.070 & 0.022 & 0.881 \\
White & 0.015 & 0.195 & 0.004 & 0.002 & 0.108 & 0.023 & 0.070 & 0.253 \\
MGQ & 0.011 & $<$0.001 & $<$0.001 & $<$0.001 & 0.009 & 0.002 & 0.537 & 0.014 \\
White, LTS screen & 0.049 & 0.046 & 0.379 & 0.075 & 0.041 & 0.050 & 0.421 & 0.041 \\
RGQ by $x_1$ & 0.006 & -- & 0.507 & 0.162 & 0.004 & 0.001 & 0.039 & 0.042 \\
\bottomrule
\end{tabular}
\end{table}

In the clean housing, savings and restaurant data, most tests detected the heteroscedasticity. After the outliers were planted, the Goldfeld--Quandt and Breusch--Pagan tests no longer detected it in the housing and savings data ($p$ between 0.36 and 0.79), and neither did White's test in the housing data ($p=0.19$). \KaHR{} still did ($p=0.009$ and $0.014$), as did MGQ. In the contaminated restaurant data \KaHR{} gave $p=0.004$. In the consumption data \KaHR{} did not reject, clean or contaminated ($p=0.19$ and $0.29$). The Breusch--Pagan test rejected only in the clean data ($p=0.022$), and with the outliers MGQ, White's test after an LTS screen and RGQ ordered by $x_1$ rejected ($p=0.014$, $0.041$ and $0.042$). The bootstrap reference, which is conservative, did not reject in any of the four contaminated data sets ($p=0.18$ to $0.55$). \KaHIII{} also rejected in the contaminated data, but for the wrong reason: its $p$-values fell because of the outliers themselves, which is the failure the robust version is designed to avoid. With parts of 6 to 10 observations, each part tolerates one or two gross errors ($b=2$ or 3), and in every data set the planted outliers fall in different parts within this limit.

The education data of the R package \pkg{robustbase} (50 US states, public education expenditure on three regressors, with Alaska as a genuine outlier) behave differently. Sorted by the first regressor, the urbanization rate, neither KaH test rejected, with or without Alaska ($p=0.96$ and 0.83 for \KaHIII{}, 0.075 and 0.11 for \KaHR{}), nor did White's test after an LTS screen ($p=0.47$ and 0.46), whereas the Breusch--Pagan ($p=0.001$) and White tests ($p=0.007$ and 0.011) rejected: the variance these score tests detect does not change along the urbanization ordering. We also used a larger data set, the 2003 Survey of Labour and Income Dynamics wages data from the R package \pkg{carData}: 4014 persons, wages regressed on education and age, sorted by education. \KaHIII{}, \KaHR{}, the Goldfeld--Quandt, Breusch--Pagan, White, MGQ and Wilcox--Keselman tests all rejected with $p<10^{-4}$. The LTS scale estimated by \KaHR{} rose from 4.8 to 5.3 to 7.0 across the thirds of education, so the variance increases with education. For log wages, the usual specification, \KaHR{} did not reject ($p=0.18$): on the log scale the variance no longer changes along education, although the Breusch--Pagan test still rejects ($p<10^{-9}$), through age.

%======================================================================
\section{Software}\label{sec:software}

The R package \pkg{KOTORY} \citep{KOTORY} implements both tests. \texttt{kah3.test()} returns the statistic $T_1$, the exact $p$-value \eqref{eq:cdf} and the critical values. \texttt{kah.robust.test()} returns $T_2$ with a choice of reference: the approximation \eqref{eq:approx} (default), a Monte Carlo reference under normal errors for any $\alpha$ and $p$, or the residual bootstrap of Section~\ref{sec:bootstrap}. \texttt{run.all.het()}, which follows the one-call test battery of \pkg{ebrahim.gof} \citep{ebrahimgof,EbrahimElKotory2026}, runs both tests next to the Goldfeld--Quandt, Breusch--Pagan, White and MGQ tests on one fitted model. The battery is descriptive: MGQ is included for comparison although its size is not controlled (Section~\ref{sec:size}), and one test should be chosen before the results are seen. The functions \texttt{pfmax()}, \texttt{dfmax()}, \texttt{qfmax()} and \texttt{rfmax()} give Hartley's distribution for any number of groups and any positive degrees of freedom. The maximum $F$ ratio distribution is also available in \pkg{SuppDists} \citep{SuppDists}; our functions evaluate \eqref{eq:cdf} by numerical integration, which allows non-integer degrees of freedom, as needed for $\nu^*$. The LTS fits use \pkg{robustbase} \citep{robustbase} and run under a fixed seed that is restored afterwards, so results are reproducible.

%======================================================================
\section{Discussion}\label{sec:discussion}

Splitting ordered data into three parts and comparing their error scales gives a heteroscedasticity test with a simple structure and a derived reference distribution. Because the parts are independent, the ratio of the largest to the smallest scale inherits a maximum $F$ law: exactly for least squares under normal errors, and with effective degrees of freedom whose limit we derived in closed form for least trimmed squares. For \KaHIII{} no simulated tables are needed; for \KaHR{} a single small table of $\nu^*$, computed once and shipped with the package, replaces the critical-value tables that a modified statistic would need for every $n$, $p$ and level.

The robust version combines three properties that no competing test combined in our study. It detects variance that changes in either direction or not monotonically at all: for U- and bulge-shaped patterns, the Goldfeld--Quandt, Breusch--Pagan, Evans--King, Wilcox--Keselman and MGQ tests were nearly powerless, and \KaHR{} was not. It can follow any ordering variable, such as time, whereas tests built on the regressors cannot see a variance that changes along a variable outside the model. And it resists outliers spread along the ordering, which destroy the level of the least squares ordered-group tests and the power of the score tests. Over a factorial design spanning these situations, its mean size-adjusted power was 67\%, against 45\% for the next test; with the variance changing along a regressor, which all tests were given, it was 67\% against 56\% for White's test after an LTS screen, and with heavy-tailed errors 64\% against 35\%. It also reports the scale of each part, which shows where in the data the variance changes.

The approach explains why two published robust tests failed. The MGQ test takes the reference distribution of a modified statistic from the unmodified one and is not valid even under normal errors. The RGQ test, implemented as described in its paper, ranks the observations by a distance that involves the response, which places the largest residuals in one group by construction. Deriving the reference of the robust statistic itself avoids the first problem, and sorting by a variable that does not depend on the errors avoids the second.

\subsection*{Limitations}

The approximation \eqref{eq:approx} is calibrated under normal errors, and with heavy tails or contamination the level of \KaHR{} rose to between 6.5 and 11\% (Fig.~\ref{fig:size}, Section~\ref{sec:factorial}, Table~\ref{tab:heavy}). The bootstrap reference keeps the level at or below 5\%, but it is conservative, and at the nominal level it loses much of the power at small $n$ (7\% against 31\% at $n=30$ with outliers, Table~\ref{tab:power}), although its size-adjusted power equals that of the default reference (Table~\ref{tab:sizeadj}); it is the choice when strict level control matters. Section~\ref{sec:robust} traces the liberal level to the normal-based $\nu^*$. On clean data the robust version pays for its trimming, and \KaHIII{} is the more powerful of the two proposed tests (Fig.~\ref{fig:factorial}b); when, in addition, the variance is monotone in a known regressor, tests built for that direction, such as the Evans--King and Goldfeld--Quandt tests, are stronger still (Table~\ref{tab:power} and Fig.~\ref{fig:power}, left panels). In the consumption data of Section~\ref{sec:data}, \KaHR{} did not detect what other robust tests detected. When outliers concentrate in one part, even within its trimming capacity, or exceed about 10\% of the data, the scale of that part is inflated and the level is lost; a global outlier screen before testing, as in White's test after an LTS screen, is then safer (Section~\ref{sec:masking}), and the reported part scales help to recognise the case. High-leverage outliers were not studied. A natural extension is to screen outliers globally before splitting the data, and to apply \KaHIII{} to the retained observations. A variance that differs between unordered groups is better detected by score tests.

%======================================================================
\section*{Computational details}\label{sec:comp}

All computations used R 4.4.1 \citep{RCore2024} with the packages \pkg{KOTORY} 0.1.0 \citep{KOTORY}, \pkg{robustbase} 0.99-4 \citep{robustbase}, \pkg{lmtest} 0.9-40 \citep{ZeileisHothorn2002}, \pkg{skedastic} 2.0.3 \citep{Farrar2020skedastic} and \pkg{carData} 3.0-5 \citep{carData}. AI-based tools were used for language polishing, for reviewing drafts and for writing code. The authors checked every result against the deposited code and output and take full responsibility for the content.

%======================================================================
\begin{appendices}

\section{Proofs}\label{app:proofs}

\paragraph{Proof of Theorem~\ref{thm:exact}.}
Under $H_0$ with normal errors, the residual vector of the least squares fit in part $P_j$ is $(I-H_j)\boldsymbol\varepsilon_j$, where $H_j$ is the hat matrix of the part and $\boldsymbol\varepsilon_j$ the error vector of the part. By Cochran's theorem, $\nu s_j^2/\sigma^2=\boldsymbol\varepsilon_j^\top(I-H_j)\boldsymbol\varepsilon_j/\sigma^2\sim\chi^2_\nu$ with $\nu=m-p-1$ when $H_j$ has rank $p+1$. Because the design is fixed and the sorting variable does not depend on the errors, the parts are fixed sets of indices, and $s_1^2,s_2^2,s_3^2$ are functions of disjoint sets of independent errors; they are therefore independent. Hence $T_1=\max_j S_j/\min_j S_j$ with $S_j=\nu s_j^2/\sigma^2$ independent $\chi^2_\nu$, which is Hartley's maximum $F$ ratio with $k=3$. For its distribution function, condition on the index $j$ and the value $x$ of the smallest $S_j$: $T_1\le c$ if and only if the other $k-1$ variables lie in $[x,cx]$. Hence
\[
P(T_1\le c)=k\int_0^\infty g_\nu(x)\{G_\nu(cx)-G_\nu(x)\}^{k-1}\,dx ,
\]
where $g_\nu$ is the $\chi^2_\nu$ density, and the substitution $u=G_\nu(x)$ gives \eqref{eq:cdf}. \qed

\paragraph{Derivation of Proposition~\ref{prop:nustar}.}
Write $Y_i=\varepsilon_i^2/\sigma^2$, independent $\chi^2_1$ with distribution function $G_1$. The statistic $\tilde\sigma^2/\sigma^2$ is the lower $\alpha$-trimmed mean $\hat\tau=h^{-1}\sum_{i\le h}Y_{(i)}$, a functional $\tau(G)=\alpha^{-1}\int_0^\alpha G^{-1}(u)\,du$ of the empirical distribution. Using the influence function of the quantile, $\{u-\mathbb{1}(y\le G^{-1}(u))\}/g\{G^{-1}(u)\}$, and integrating over $u\in(0,\alpha)$ with the substitution $t=G^{-1}(u)$, the influence function of $\tau$ is
\[
\operatorname{IF}(y)=\xi-\tau-\alpha^{-1}(\xi-y)_+ ,\qquad \xi=G_1^{-1}(\alpha),
\]
so that $m\operatorname{var}(\hat\tau)\to\operatorname{var}\{(\xi-Y)_+\}/\alpha^2$ \citep{Stigler1973}. By the delta method, $m\operatorname{var}(\log\hat\tau)\to\operatorname{var}\{(\xi-Y)_+\}/(\alpha^2\tau^2)$. For a $\chi^2_{\nu^*}/\nu^*$ variable, $\operatorname{var}\{\log(\chi^2_{\nu^*}/\nu^*)\}=\psi'(\nu^*/2)=2/\nu^*+O(\nu^{*-2})$. Equating the two variances, and using the assumed asymptotic equivalence of the squared LTS scale with a constant multiple of $\tilde\sigma^2$, which leaves the variance of the logarithm unchanged, gives $\nu^*/m\to r(\alpha)$ in \eqref{eq:ralpha}. Since $\nu=m-p-1$ with $p$ fixed, $\nu^*/\nu$ has the same limit. As $\alpha\to1$, $\xi\to\infty$, $\operatorname{var}\{(\xi-Y)_+\}\to\operatorname{var}(Y)=2$ and $\tau\to1$, so $r(\alpha)\to1$. \qed

\end{appendices}

%======================================================================
\backmatter

\section*{Statements and Declarations}

\noindent\textbf{Funding.} No funding was received for conducting this study.

\smallskip\noindent\textbf{Competing interests.} The authors have no relevant financial or non-financial interests to disclose.

\smallskip\noindent\textbf{Ethics approval.} Not applicable.

\smallskip\noindent\textbf{Consent.} Not applicable.

\smallskip\noindent\textbf{Data availability.} The wages and education data are in the R packages \pkg{carData} and \pkg{robustbase}; the four benchmark data sets of Section~\ref{sec:data}, with the planted outliers, are in the Zenodo archive. The simulation code, the per-replication $p$-values and the scripts that produce every table and figure are archived at Zenodo, \url{https://doi.org/10.5281/zenodo.23038198} (this DOI resolves to the latest version).

\smallskip\noindent\textbf{Code availability.} The R package \pkg{KOTORY} is available at \url{https://github.com/ebrahimkhaled/KOTORY} (submitted to CRAN); the version used in this paper, 0.1.0, is included in the Zenodo archive above.

\smallskip\noindent\textbf{Author contributions.} A.E.-K. proposed the tests. E.K.E. derived the null distributions and the effective degrees of freedom, designed and ran the simulation study, wrote the software and drafted the manuscript. Both authors read and approved the final manuscript.

\bibliography{references}

\end{document}